# Multi-Objective Particle Swarm Optimization for Facility Location Problem in Wireless Mesh Networks

Tarik Mountassir[1], Bouchaib Nassereddine[1,3], Abdelkrim Haqiq[1,2] and Samir Bennani[3]

[1] IR2M Laboratory, FST, Hassan 1st University, Settat, Morocco

[2] e-NGN research group, Africa and Middle East

[3] RIME Laboratory, Mohammedia Engineering School Agdal, Rabat Morocco

**Abstract**

Wireless mesh networks have seen a real progress due of their implementation at a low cost. They present one of Next Generation Networks technologies and can serve as home, companies and universities networks. In this paper, we propose and discuss a new multi-objective model for nodes deployment optimization in Multi-Radio Multi-Channel Wireless Mesh Networks. We exploit the trade-off between network cost and the overall network performance. This optimization problem is solved simultaneously by using a meta-heuristic method that returns a non-dominated set of near optimal solutions. A comparative study was driven to evaluate the efficiency of the proposed model.

## 1. Introduction

Wireless mesh networks (WMN) have emerged as a reliable and cost efficient for providing large coverage area through mutli-hop wireless communications and for improving wireless ad-hoc Networks, Local Area Networks (LAN), Personal Area Networks (WPAN) and Metropolitan Area Networks (WMAN). Thus, the mesh networking concept can be implemented by IEEE 802.11, 802.15 and 802.16 technologies. In WMNs, Wireless connection between a sender and a receiver can be ensured by redundant paths. This increase the network reliability and robustness.

Such networks consist of Mesh Clients (MCs) and three types of interconnected wireless routers; Fig.1 illustrates a relatively static infrastructure of a WMN composed by: Access Points (APs) which provide coverage to MCs, Mesh relays (MR) which route traffic to other MRs/APs and Mesh Gateways (MGs) which have an Internet connection [1]. These routers are organized as the wireless network backbone and can be equipped by multiple wireless interfaces. Furthermore, the use of Multiple-Radio Multiple-Channel techniques improves the capacity of mesh networking but it result more interferences. Thus the planning of such networks presents many challenges for network operators.

Several applications for broadband wireless services have been deployed based on WMNs. They include community and neighborhood networks, public safety and security, building and electric utility automation and transportation systems. However, these applications continue to confront the problem of connectivity and performance by poor planning of wireless networks. Recently, considerable interest has been given to WMNs design problem. Most of these studies have focused on routing [2], interference measurement and capacity analysis [3], power control [4], topology control [5], link scheduling [6], channel/radio assignment [7] and nodes placement [8].

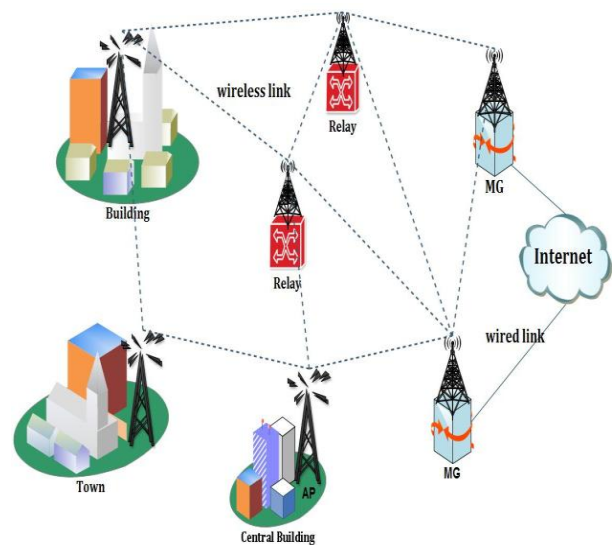

Fig. 1 Architecture of a WMN

However, few studies have addressed the nodes placement and the existing contributions have not considered all the parameters that influence the quality of results.

WMNs planning can be divided into two stages which are conducted either simultaneously or separately. First, nodes deployment is a central problem which consists in positioning MRs in an area in order to provide a network access with the desired coverage and Qos requirements while minimizing deployment cost. This issue corresponds to the Facility Location Problem and the Set Covering Problem [2] which is NP-hard problem [3]. Second, the channel assignment is also a NP-hard problem that aims to assign frequencies to each radio interface in order de minimize interference. This problem is known as Automatic Frequency Planning.

The network cost, namely the number of routers, and the overall network performance present a trade-off that make WMN planning a sensitive task. This is an open research issues that need to be investigated.

Many algorithms have been proposed for WMN planning. The authors of [10], [11], [12] optimized the placement of MRs (i.e. the backbone WMNs) to provide large coverage to clients at a minimum cost while guarantying good performances and a minimum level of interference.

Other contributions (eg [13], [14], [15]) have introduced different methods to select a minimum number of MRs to become gateways while satisfying the throughput, interference and congestion constraints. In [16], the authors define a generalized form of a linear program that takes into account interference and transmission power to minimize the cost function. The study of [17] presents an optimization model for WMNs planning that aims to minimize the network deployment cost while providing complete clients coverage. Most studies proposed models with one objective which is to minimize the cost and opted for exact optimization techniques (CPLEX for example) to find optimal planning solution. However, the planning problem invokes multiple conflicting performance measurements or objectives to be simultaneously optimized. In this context, a multi-objective formulation of this problem has been proposed in [18] where the authors simultaneously optimized the cost, the level of interference on all network links and minimized gateways neighborhood congestion.

In recent paper [19], we considered the Multi-Radio Multi-Channel WMNs planning where we proposed a multi objective model for nodes deployment problem that were based on maximizing total clients coverage and links load balancing while minimizing deployment cost.

We also proposed, in [20], a relays placement algorithm that provide newtork connectivity and robustness while in [21], we proposed and disscussed tow new models for nodes deployment that, beside minimizing cost and maximizing total clients coverage, attempted to balance both links and gateways loads as the common objective functions subject to the same group of constraints. This permited us to evoke a comparative study of the three models. The optimization problems was solved by using meta-heuristic method that return non-dominated set of near optimal solutions.

In this contribution,in the one hand, we propose a new and powerful model that combines four conflictuel objectives for the facility location problem in wireless mesh networks including: Cost minimization, maximization coverage, links, congestion minimization and gateways-congestion minimization. On the other hand, we use a meta-heuristic method to resolve this problem, being NP-hard, and we propose an efficient algorithm for nodes deployment problem. A comparative study, between different other model, will be conducted to show the efficiency of our complex model.

To the best of out knowledge, this is the first contribution that proposes a model associating all these conflictuel objectives functions and that proceed to a simultanous multi-objective approach to resolve this problem known as the facility location problem.

The rest of the paper is organized as follow: Network model and problem formulation are presented in Section II. In Section III, we propose a multi-objective optimization of the proposed models and we present numerical results and analysis. We conclude our work and give directions for future research in Section IV.

## 2. Network Model and Formulation

### 2.1 Network Model

We consider a Multi-Channel Multi-Radio WMN represented by a graph $G = (V, E)$ where $V$ is the set of wireless routers and $E$ describes the set of links between each pair of MRs. We assume the MRs have the same number of radio interfaces $R$. Each one is equipped with $K$ channels ($K > R$). A link can be established between two MRs when one of their interfaces use the same channel and the distance between them is less than the transmission range of each MR. To consider interference between the links of WMNs, we adopt the Protocol Interference Model [3] where a transmission on channel $k$ is successful when all interferes in the neighborhood of the transmitter and receiver are silent during the transmission time. Let $N = \{1,$

..., n} be the set of traffic Demand Points (Mesh Clients) and $S = \{1, ..., s\}$ be the set of Candidate Sites to host a node (AP, MR, MG). Table 1 shows the notations used to describe the model.

Table 1: List of parameters and variables

| | |
|---|---|
| $n$ | Number of Demand Points (DP) |
| $s$ | Number of Candidate Sites (CS) |
| $T_i$ | Traffic Demand of DP $i$ |
| $C^k_{jl}$ | Capacity of link $(j,l)$ using channel $k$ |
| $C_{max}$ | Maximum capacity of the radio interface of a router |
| $R$ | Number of radio interfaces per node |
| $K$ | Number of channels per radio interface |
| $a_{ij}$ | Coverage of a DP $i$ by CS $j$ |
| $b_{jl}$ | Radio connectivity between two candidate sites $j$ and $l$ |
| $z_j$ | Installation of a nodes on CS $j$ |
| $n_j$ | Installation of an Access Point at CS $j$ |
| $r_j$ | Installation of a Router at CS $j$ |
| $g_j$ | Selection of a Gateway to CS $j$ |
| $x_{ij}$ | Assignment of DP $i$ to CS $j$ |
| $w_j^k$ | Installation of a router at CS $j$ using the channel $k$ |
| $L_{jl}^k$ | Establishing radio communication between CSs $j$ and $l$ using the channel $k$ |
| $f_{jl}^k$ | Flow on channel $k$ between CSs $j$ and $l$ |
| $F_j$ | Flow between the gateway and the ISP |
| $Q$ | All frequency channels $Q = \{1, ..., K\}$ |
| $A$ | Constant |

We consider the following binary parameters namely:

$$a_{ij} = \begin{cases} 1 \text{ if } DP\ i \text{ is covered by } CS\ j \\ 0 \text{ otherwise} \end{cases} \quad (1)$$

$$b_{ij} = \begin{cases} 1 \text{ if two } CS\ (j,l) \text{ are wiressely} \\ \qquad\qquad connected \\ 0 \text{ otherwise} \end{cases} \quad (2)$$

$$z_j = \begin{cases} 1 \text{ if a node is installed on } CS\ j \\ 0 \text{ otherwise} \end{cases} \quad (3)$$

$$n_j = \begin{cases} 1 \text{ if an AP is installed on } CS\ j \\ 0 \text{ otherwise} \end{cases} \quad (4)$$

$$r_j = \begin{cases} 1 \text{ if a Mesh Relay is installed on } CS\ j \\ 0 \text{ otherwise} \end{cases} \quad (5)$$

$$g_j = \begin{cases} 1 \text{ if a MG is installed on } CS\ j \\ 0 \text{ otherwise} \end{cases} \quad (6)$$

$$x_{ij} = \begin{cases} 1 \text{ if a } DP\ i \text{ is assigned } CS\ j \\ 0 \text{ otherwise} \end{cases} \quad (7)$$

$$w_j^k = \begin{cases} 1 \text{ if device on } CS\ j \text{ using channel } k \\ 0 \text{ otherwise} \end{cases} \quad (8)$$

$$L_{jl}^k = \begin{cases} 1 \text{ if a wireless link is established} \\ \text{between } CSs\ j \text{ and } l \text{ using channel } k \\ 0 \qquad otherwise \end{cases} \quad (9)$$

In this paper we consider all the parameters that influence the quality of the solution. Thus, we have formulated our model through a multitude of criteria that have a significant impact on the result and we use Integer Linear Programming to express the Facility Location Problem. These criteria include constraints to be satisfied and four objective functions to optimize.

2.2 Objective Functions

Most of the proposed networks planning solutions consider the cost as the only function to optimize. However, the problem invokes many performance measurements and objectives to be simultaneously optimized. This approach exploits the trade-off between conflictual functions and provides more realistic solution.

In this contribution, we propose an integrated model that we call *Link-Gateways Load Balancing* (**LGLB**) that contains four objective functions to be simultaneously optimized namely:

**Cost function:**

$$Min \sum_{j \in S} (n_j + r_j + g_j) \quad (10)$$

This function minimizes the total deployment cost by positioning a minimum number of APs, MRs and MGs.

**Coverage function:**

$$Max \sum_{i \in N} \sum_{j \in S} a_{ij}\ r_j \quad (11)$$

The objective is to maximize the number of clients to be covered by installed routers.

**Links load balancing:**

As the flow on a link approaches to its capacity, the link becomes more prone to congestion. The objective is to maximize the number of links that has minimum congestion by balancing the traffic load among different links. Thus, We put:

$$Max \left( Min_{\substack{j,l \in S, \\ k \in Q}} \left( L_{jl}^k C_{jl}^k - f_{jl}^k \right) \right) \qquad (12)$$

**Gateways load balancing:**

The objective is to minimize congestion around gateways by minimizing the unfair use of MG measured by the standard deviation of flows entering network through gateways. This function was used in [18]:

$$Min \sqrt{\frac{\sum_{l \in S} F_l^2}{\sum_{l \in S} F_l}} \qquad (13)$$

2.3 Problem Constraints

The problem constraints are formulated as follow:

**Coverage constraint:**

The problem must make sure that a given DP $i$ is assigned to at most one CS $j$. Thus:

$$\sum_{j \in S} x_{ij} \le 1 \qquad \forall i \in N \qquad (11)$$

A given DP $i$ must be affected and covered by an installed node in CS $j$. Thus:

$$x_{ij} \le a_{ij} z_j \qquad \forall i \in N, \forall j \in S \qquad (12)$$

**Links - Interference constraints:**

A node can use at most R radio interfaces for transmission or reception or both. Thus:

$$\sum_{j \in S} \sum_{k \in Q} L_{jl}^k \le R \qquad \forall l \in S \qquad (13)$$

The maximum number of channel that can be used on link $(j,l)$ is equal to $K$. Thus:

$$\sum_{k \in Q} L_{jl}^k \le K \qquad \forall j, l \in S \qquad (14)$$

To prevent a mesh node from selecting the same channel $k$ more than once to assign it to its interfaces, we put:

$$\sum_{l \in S} L_{jl}^k \le 1 \qquad \forall j \in S, \forall k \in Q \qquad (15)$$

To avoid simultaneous transmission or reception using the same channel, we have:

$$\sum_{j \in S} L_{jl}^k + \sum_{l \in S} L_{jl}^k \le 1 \quad \forall k \in Q \qquad (16)$$

A link between CS $j$ and CS $l$ can exist only when the two devices are installed, wirelessly connected and tuned to the same channel $k$. thus:

$$2 L_{jl}^k \le b_{jl} \left( w_j^k + w_l^k \right) \quad \forall j, l \in S, \forall k \in Q \qquad (17)$$

We state that the number of links from a mesh node is limited by the number of radio interfaces. Thus:

$$\sum_{k \in Q} w_j^k \le R z_j \qquad \forall j \in S \qquad (18)$$

**Flow – Capacity constraints:**

The sum of requested service by node on CS $j$ must not exceed the capacity of radio interface. Thus:

$$\sum_{i \in N} T_i x_{ij} \le C_{max} \qquad \forall j \in S \qquad (19)$$

The flow on a link cannot exceed the capacity of this link. This can be expressed by:

$$f_{jl}^k \le L_{jl}^k C_{jl}^k \qquad \forall j, l \in S, \forall k \in Q \qquad (20)$$

To define the network flow balance, we put:

$$\sum_{i \in N} T_i x_{ij} + \sum_{j \in S} \sum_{l \in S} \left( f_{jl}^k + f_{lj}^k \right) = F_j \quad \forall j \in S \qquad (21)$$

If a mesh node is installed on CS $j$ as a gateway and AP is installed in CS $l$, then the path length (hops number) between $j$ and $l$ cannot exceed $A$ hops:

$$2 h_{jl} \le A (z_j + z_l) \qquad (22)$$

A node installed on CS *j* can route the flow to the Internet only if it is a gateway. This flow is limited by a large number M. Thus:

$$F_j \leq M g_j \qquad \forall j \in S \qquad (23)$$

**Robustness constraint:**

Once node is deployed at a CS *j*, it is required that there are at least two nodes in disjoint paths connecting it to the network. This ensures any single failure does not disconnect the network. We put:

$$\sum_{j \in S} \sum_{k \in Q} L_{jl}^k \geq 2 \qquad \forall l \in S \qquad (24)$$

All parameters considered in the formulation must respect the following conditions:

$$r_j, n_j, z_j, g_j, x_{ij}, L_{jl}^k, w_j^k \in \{0, 1\} \qquad (25)$$
$$\forall i \in N, \forall j \in S$$

$$f_{jl}^k, F_j, M \in R^+ \qquad \forall j, l \in S \qquad (26)$$

## 3. Results and Analysis

In this section, we propose and analyze solutions for nodes placement models presented in previous section.

### 3.1 Problem Resolution

We use the evolutionary technique called Multi-Objective Particle Swarm Optimization (MOPSO) proposed in [20] where authors introduced the mechanism of Crowding Distance to maintain diversity in the Pareto frontier. This method is based on the PSO of Kennedy and Eberhart [21] which is built on the social behavior of flocks of birds that tend to imitate successful actions they see around them, while there bringing their personal variations. A swarm (population) consists of several particles (individuals).

In this method, the building of an initial solution that meets the constraints is a sensitive task that should be done carefully. This leads us to propose the algorithm 1 where we first randomly generate a coverage matrix that expresses the allocation of every $DP_i$ to a $CS_j$ and another connectivity matrix between each CSs *j* and *l*. Both matrixes contribute to the construction of initial feasible solutions, representing the placement of Mesh APs, Mesh Relays and MGs, which will be stored in the archive. A particle represents the set of binary variables representing the solution of the problem namely: $n_j$, $r_j$, $g_j$, $L_{jl}^k$ and $f_{jl}^k$.

We considered the planning of a network with a grid topology and we proceeded to the placement of nodes in three steps:

1. *Access Points Placement:* We placed randomly the nodes in CSs to cover all the DPs that are not yet assigned to an access point.

2. *Relays Placement*: We added new routers to connect the graph G (V, E) by requiring all nodes to have at least two neighbors which ensure robustness to the network ( eg. Algorithm 2).

3. *Gateways Placement*: We choose randomly among the nodes deployed in steps 1 and 2, who will act as a gateway.

---

**Algorithm 1**: Topology Auto Planning

**Input**: *mut*: mutation factor, *gmax*
**Output**: Archive
**Begin**
Initialize swarm    //Construct initial feasible solutions
Evaluate all particles in swarm  // Compute Objectives
Store all non-dominated solutions into the Archive
**Repeat** (g < gmax)
   **For** each particle in the archive
     Compute Crowding Distance (CD) value,
     Sort the Archive in a descending order of CD values
     Mutation (mut)
     ***Access Points Placement***
     ***Relays Placement*** // invoke Algorithm 2
     ***Gateways Selection***
  **EndFor**
     Check for constraints satisfaction
     Evaluation   // Compute Objective functions
     Update Archive
     g++
**Until** ( g >= gmax )
**End**

---

Furthermore, we check for constraints satisfaction and evaluate the objective functions at every generation of the swarm in order to provide different non-dominated solutions.

```
Algorithm 2: Relays Placement
Input:   Set of APs
Output   Set of Mesh Relays
Begin
Initialize r   //MRs number   r = 0
Repeat

   For each AP j in A
      If  (AP j is placed on a corner)
         Add 2 nodes in the neighborhood  //  r = r + 2
      End if
      If  (AP j is placed on an edge)
         Add 3 nodes in the neighborhood //   r = r + 3
      End if
      If  (AP j is placed on an internal)
         Add 4 nodes in the neighborhood //   r = r + 4
      End if
   End For

Until ( all APs are visited )

End
```

3.2 Numerical Results and Analysis

To show the quality of the proposed solution, we study the performance of the models and we consider 6×6, 7×7, 8×8 and 10×10 grid topologies as CSs where mesh nodes can be installed. We define standard settings by 200 demand points (groups of clients), the traffic demand $T_i$ = 2Mb/s, $C_{jl}^{k}=C_{max}$ =54Mb/s, $A = R = 3$ and number of channel $k$ = 11. The algorithm was coded in JAVA language and all experiments were carried out on a Core i3 machine.

It is clear that one of the most important goals of WMNs planning is cost minimization (represented by mesh nodes numbers in our model). Thus, we run our algorithm and we selected only the cheapest solutions which are located in the Pareto frontiers.

By varying the number of CSs, we observe in Fig. 2 that the number of nodes increases when the number of CSs increases. This is caused by the need of connecting the network and providing more bandwidth respectively.

We also studied the effect of changing Ti on deployment cost showed in Fig. 3 while maintaining standard parameters fixed. When Ti increases, we see the number of APs and gateways increases highly to satisfy this service demand to cover more clients. However, it result a low fairness on links so the throughput decreases.

Finally, an optimal number of radio interfaces permits the maximization of throughput and the demand satisfaction with minimum number of nodes. Fig. 4 and show that when R = 2, the coverage and the throughput are maximized. While the cost is minimized when R= 3.

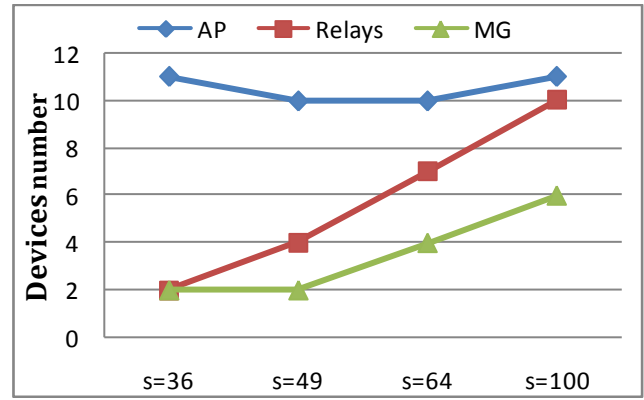

Fig. 2 Number of nodes when S varies

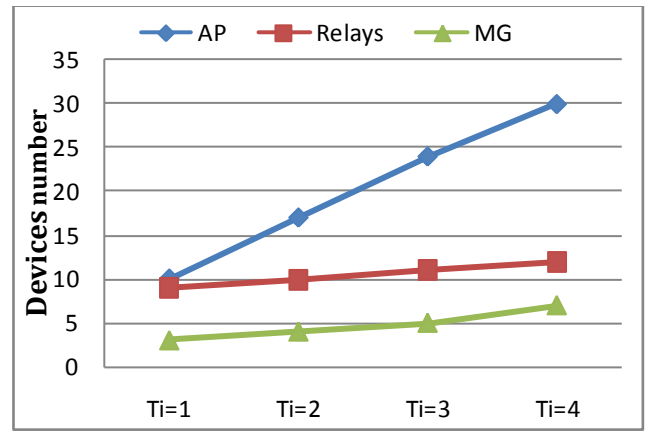

Fig. 3 Number of nodes when Ti varies

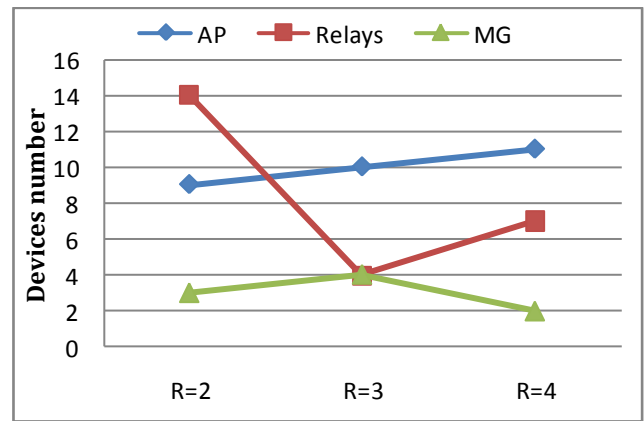

Fig. 4 Number of nodes when R varies

## 3.3 Comparative study

In the following, we proceed to a comparative study of our integral LGLB model with other models already proposed and studied by the authors of [21] ie Coverage model (COV), Link Load Balancing model (LLB) and Gateways Load Balancing model (GLB). We run our algorithm for the four models and we selected only the cheapest solutions which are located in the Pareto frontiers. Thus, we plotted for each model nodes number when CSs number varies and all other parameters are fixed. The planning solution will be provided for 200 Demand Points for all models.

The comparison between the four models is made by analyzing the cost function represented by the number of APs (Plotted in Fig 5), Relays (Plotted in Fig 6), and MGs (Plotted in Fig 7).

Fig.5 shows that the number of APs provided by the Coverage Model (COV) is greater than that given by the other three models. This can be explained by the fact that function (11) maintains client coverage as maximal as possible while satisfying function (10).

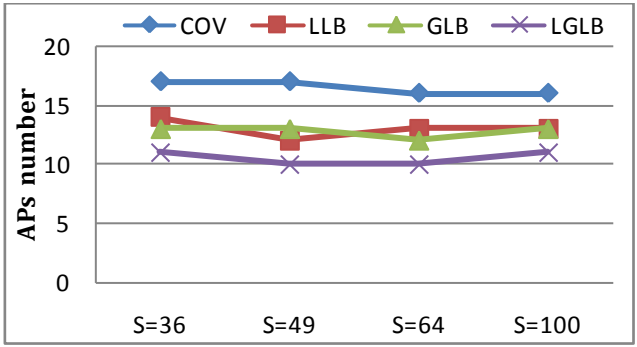

Fig. 5 APs Number

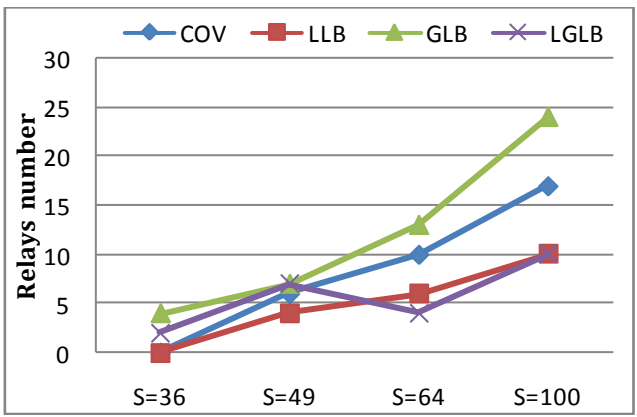

Fig. 6 Relays number

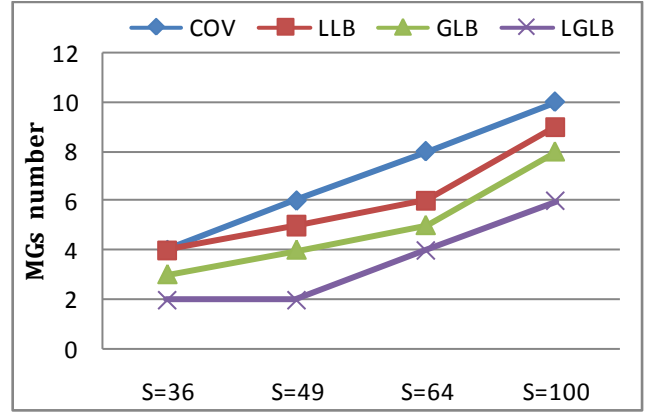

Fig. 7 MGs number

Adding function (12) LLB model showed that the algorithm generates less relays. Thus, in Fig.6, LLB and LGLB models seem to be better than other models.

When we add a gateway throughput improvement function in GLB model, we observe that a maximum throughput exploited by all users while satisfying the economic criteria (Fig.7).

Finally, we note that the model proposed by this article namely LGLB model, which contains four functions optimized simultaneously provides a complete and effective solution for Facility Location problem in WMNs.

## 4. Conclusions

In this paper, we proposed a new integrated mutli-objective model for wireless mesh networks planning by optimizing four objective functions simultaneously subject to a set of constraints to take into account namely interference, robustness and load balancing. The use of the Multi-Objective Particle Swarm Optimization method to resolve our models provides very interesting results and lets the network planner decide which solution responds to his requirements. Additionally, we compared the performance of our proposed model with other previous models by comparing the algorithm solutions in term of deployment cost expressed as access Points, Relays and gateways number.

As further research topic, we intend to propose a hybrid algorithm for WMNs Planning that aims to simultaneously place wireless nodes and allocate channels/links.

**Tarik MOUNTASSIR** obtained the Master of Science and Technology in electronics and the Higher Depth Studies Diploma in computer sciences from CADY AYYAD University, Morocco, in 2005 and 2007 respectively where he was major his class. He worked as project manager engineer for the central popular bank in morocco from 2008 to 2010 and today he is an IT manager at the Ministry of Interior of Morocco. Mr Mountassir Is now a PhD student in computer science at HASSAN 1st University, Morocco.

**Bouchaib Nassereddine** is a Professor in Faculty of sciences and Techniques, Hassan 1rst University, SETTAT Morocco. He is an active member of research team RIME in Ecole Mohammadia d'Ingénieurs - Rabat and Computer, Networks, Mobility, and Modeling Laboratory, Hassan 1st University SETTAT.

**Abdelkrim Haqiq** is a member of the IEEE and the IEEE Computer Society.

**Samir Bennani** is a Professor and Computer Science department Head in Ecole Mohammadia d'Ingénieurs - Rabat. He is the research team RIME (Networking, Modeling and eLearning) director.